\documentclass[preprint,12pt]{elsarticle}
\usepackage{epsfig}
\usepackage{amssymb}
\usepackage{slashed}
\usepackage{amsmath}
\usepackage{cancel}
\usepackage{float}
\usepackage{color}
\journal{Nuclear Physics A}

\newcommand{\ga}{\alpha}
\newcommand{\gb}{\beta}

\newcommand{\be}{\begin{equation}}
\newcommand{\ee}{\end{equation}}
\newcommand{\ba}{\begin{eqnarray}}
\newcommand{\ea}{\end{eqnarray}}
\newcommand{\dd}{{\rm d}}

\begin{document}

\begin{frontmatter}

\title{Entropy production for an interacting quark-gluon plasma}

\author{Stefano Mattiello}
\address{Institute for Theoretical Physics, University of Gie{\ss}en, Germany}


\begin{abstract}
We investigate the entropy production within dissipative
hydrodynamics in the Israel-Stewart (IS) and Navier-Stokes theory (NS) 
for relativistic heavy ion physics 
applications. In particular we focus on the initial
condition in a 0+1D Bjorken scenario, appropriate
for the early longitudinal expansion stage of the collision.
Going beyond the standard simplification of a massless ideal gas we consider
a realistic equation of state consistently derived within a virial
expansion. The EoS used is well in line with recent
three-flavor QCD lattice data for the
pressure, speed of sound, and interaction measure at nonzero
temperature and vanishing chemical potential ($\mu_{\rm q} = 0$).
The shear viscosity has been consistently calculated within this
formalism using a kinetic approach in the ultra-relativistic regime with an
explicit and systematic evaluation of the transport cross section as function
of temperature.
We investigate the influence of the viscosity and the initial condition,
i.e. formation time, initial temperature, and pressure anisotropy for the
entropy production at RHIC at $\sqrt{s_{\rm NN}}=130$ GeV.
We find that the interplay between effects of the viscosity and of the
realistic EoS can not be neglected in the reconstruction of the initial
state from experimental data.
Therefore, from the experimental findings it is very hard to
derive unambiguous information about the initial conditions and/or the
evolution of the system.
\end{abstract}


\end{frontmatter}

\section{Introduction}

Understanding the rich phase structure of quantum chromodynamics for
the density-temperature plane is a challenge for theoretical as well as
expe\-rimental high energy physics.
Lattice Monte Carlo simulations have revealed several exciting results over the
past
decade~\cite{Tannenbaum:2007dx,Fodor:2002km,Fodor:2001pe,Allton:2003vx,Allton:2005gk,D'Elia:2002gd,D'Elia:2004at,Cheng:2007jq}.
At high densities, where the lattice calculations can be
performed, effective models of QCD are needed for a theoretical
description of this rich phase
structure
~\cite{Alford:1997zt,Alford:1998mk,Rapp:1997zu,Rajagopal:2000wf,Klevansky:1992qe,Bluhm:2004xn,Bluhm:2007nu,Ratti:2005jh,Cassing:2007yg,Cassing:2007nb,Cassing:2008nn,Mattiello:2004rd,Strauss:2009uj,Mattiello:2009fk}.
From the experimental point of view, heavy-ion collisions are the tool for such investigations.
Recent observations at the relativistic heavy-ion collider (RHIC) at Brookhaven National
Laboratory (BNL) indicated that the quark-gluon plasma (QGP) created in
ultra-relativistic Au + Au collisions is a strongly interacting system.
Due to this experimental evidence the QGP can not be satisfactory described by the
Stefan-Boltzmann (SB) limit for relativistic noninteracting massless particles, but a
realistic equation of state (EoS) has to be used.
Recently, we have systematically derived such an EoS within a virial
expansion~\cite{Mattiello:2009fk}. The recent three-flavor lattice QCD data are described very
well for the main thermodynamics
quantities, e.g., pressure, entropy density, speed of sound, and interaction
measure.
Regarding heavy ion collisions, we note that their hydrodynamical modeling
plays a crucial role in deducing the experimental findings~\cite{Huovinen:2006jp}.
Important sources of uncertainty are given by the initial conditions, that are
not known precisely.
The experimental estimation of the initial energy density has to be taken with
care, because it contains simplifications.
In fact, Bjorken estimation of the initial energy density at a
conservative thermalization time $\tau_0=1 {\rm fm/c}$ is used assuming a non
interacting QGP (i.e. SB equation of state) and a transverse energy
distribution per unit of rapidity proper time independent during the evolution of the
system~\cite{Arsene:2004fa,Back:2004je,Adams:2005dq,Adcox:2004mh}.
This second assumption allows to avoid any estimation of the lifetime
of the deconfined phase but is a strong hypothesis that can lead to changes
in the values of the transverse energy density by a factor 2 at times between
1 and 8 fm/c~\cite{Pal:2003rz}.
Therefore, already in the assumption of a non dissipative QGP the estimation of
the initial energy density has to be considered with care.
The importance of the initial condition has been shown by the
large elliptic flow, $v_2$, in
Au+Au and Cu+Cu collisions at RHIC  
energies~\cite{Adler:2001nb,Adler:2002pu,Adams:2003am,Adams:2004bi,:2008ed,Adcox:2002ms,Adler:2003kt,Adler:2004cj,Adare:2006ti,Back:2002gz,Back:2004zg,Alver:2006wh}.
The centrality, pseudorapidity, and transverse momentum
dependences of $v_2$ data are described reasonably well
by employing the Glauber-type initial conditions and
implementing hadronic dissipative effects
in ideal hydrodynamic models~\cite{Hirano:2005xf}.
However, by replacing the initial conditions from the Glauber
model to the ones expected from a color glass condensate,
elliptic flow coefficients overshoot the experimental data.
This is due to eccentricity larger than the ones in the conventional Glauber model.
This discrepancy strongly suggests effects of viscosity in the
QGP~\cite{Monnai:2009ad}. 
A combined investigation of both effects, i.e., viscosity/anisotropy
effects and the dependence on the initial condition is needed to understand
the competition between these two aspects thoroughly.
A measurable quantity that allows such an investigation is the entropy density
per unit of rapidity ${\rm d}S/{\rm d}y$.
In absence of viscous effects ${\rm d}S/{\rm d}y$ is a conserved
quantity and therefore delivers direct information about the initial
condition of the system~\cite{Pal:2003rz}, depending on the EoS used.
On the other hand, if the shear viscosity, $\eta_{\rm s}$, can not be neglected, ${\rm d}S/{\rm d}y$
increases, and the details of the dynamical evolution become important.
Hydrodynamical calculations use a simplified picture for $\eta_{\rm s}$ and for the
equation of state.
Usually the viscosity to entropy density ratio as temperature independent
quantity~\cite{Monnai:2009ad,Luzum:2008cw,Song:2008si,Dusling:2007gi,Schenke:2010nt}
and/or noninteracting SB-EoS are
used~\cite{Huovinen:2008te,Molnar:2009pq,Martinez:2009mf,Martinez:2009ry}.
Only recently, calculations including a schematic temperature dependence of $\eta_{\rm s}/s$
in the hadronic phase~\cite{Shen:2011kn} as well as in the
QGP~\cite{Niemi:2011ix} have been performed.

Therefore, a consistent description, where a realistic equation of state as well
as the temperature dependent shear-viscosity derived within the same approach
is needed to systematically investigate the competition between
viscosity and initial condition effects on the hadronic yields.
Such a consistent approach is given by the virial expansion, since it allows to
derive not only a realistic equation of state for the QGP but also to resolve the whole temperature dependence of the shear viscosity within a
kinetic theory~\cite{Mattiello:2009db}.
Starting from the results of our model, in this work we will focus on the investigation of the role of
the viscosity and the initial conditions for the evolution and finally for
the entropy production in heavy-ion collisions.
In order to understand the interplay of viscosity, realistic EoS, and initial
conditions we consider a $0+1$-dimensional time evolution.
This simplified scenario is appropriate
for the early longitudinal expansion stage of the collision.
Because we focus on the entropy production, which occurs mostly during
the early stage of the expansion~\cite{Song:2008si}, this approximation is completely justified.
In fact, the authors of Ref.~\cite{Song:2008si} claim, that, for a massless ideal gas, the
entropy production can be calculated to excellent approximation by assuming
boost-invariant longitudinal expansion without transverse flow during this period.

The present work is organized as follows: In
Section~\ref{Sec:Theorie} we briefly recall the basics of the
hydrodynamical equation of motion for a shear-viscous, longitudinally boost and
transverse translation invariant system.
In particular, we discuss how the relativistic perfect fluid (Euler) and the Navier-Stokes
equations of motion can be derived from these equations as a special case.
In Section~\ref{Sec:EntrProd} we give a brief
explanation  of the 
entropy per unit of rapidity in the different scenarios.
In Section~\ref{Sec:Results} we present our results by
discussing in detail the role of the viscosity and the initial condition,
i.e., formation time, initial temperature, and pressure anisotropy for the
evolution of the system and the entropy production at RHIC at $\sqrt{s_{\rm
    NN}}=130$ GeV and the consequences for an improvement of their experimental
estimation.
The conclusions in Section~\ref{Sec:Concl}
finalize this work.

\section{Boost-invariant hydrodynamics}\label{Sec:Theorie}
Relativistic hydrodynamic is based on the local energy-momentum 
and charge conservation
\be
\partial_\mu T^{\mu\nu}(x) = 0 \quad , \qquad \partial_\mu n^\mu(x) = 0 \ ,
\label{eqn:hydro}
\ee
expressed in terms of the energy-momentum tensor $T^{\mu\nu}(x)$ and of the
local charge density $n(x)$.
In ideal (Euler) hydrodynamics dissipative effects are neglected. Therefore the
energy-momentum tensor and the charge density are given by
\ba
T^{\mu\nu}_{\rm ID}&=& (\varepsilon+P)u^\mu u^\nu-Pg^{\mu\nu}\\
n^{\mu}_{\rm ID}&=& n u^\mu,
\ea
where $\varepsilon$ and $P$ are the energy density and the pressure,
respectively; $u^\mu$ is the flow four-velocity normalized to $u^\mu u_\mu=1$
using the standard metric tensor $g^{\mu\nu}={\rm diag}(1,-1,-1,-1)$.
The simplest extension to a dissipative regime is the introduction of additive
corrections linear in flow and temperature gradients~\cite{Huovinen:2008te}, the so-called
Navier-Stokes (NS) approximation
\ba
\delta T^{\mu\nu}_{\rm NS} &=&
\eta_{\rm s} (\nabla^\mu u^\nu + \nabla^\nu u^\mu 
        - \frac{2}{3}\Delta ^{\mu\nu} \partial^\alpha u_\alpha) 
+ \zeta \Delta^{\mu\nu} \partial^\alpha u_\alpha, \quad
\label{eq:NS_T}
\\
\delta n^\mu_{\rm NS} &=& 
\kappa_{\rm q} \left(\frac{n T}{\varepsilon+p}\right)^2 
\nabla^\mu \left(\frac{\mu}{T}\right), \quad
\label{eq:NS_n}\\
\Delta^{\mu\nu} &\equiv& g^{\mu\nu} - u^\mu u^\nu \qquad {\rm and}\qquad
\Delta^\mu \equiv \Delta^{\mu\nu} \partial_\nu, 
\ea
where $\zeta$ indicates the bulk viscosity, and $\kappa_{\rm q}$ is the heat
conductivity of the system.
The hydrodynamic equations in NS approximation  
can be derived from a general non-equilibrium theory,
the on-shell covariant transport~\cite{Gyulassy:1997ib,Molnar:2001ux}.
The relativistic NS equations 
are parabolic and therefore acausal.
The solution formulated by Israel and Stewart (IS)~\cite{Israel:1976tn,Israel:1979wp} transforms the NS equations
into relaxation equations for the shear stress,
$\pi^{\mu\nu}$, bulk pressure, $\Pi$, and heat flow, $q^\mu$.
The corrections are given by
\be
\delta T^{\mu\nu} \equiv \pi^{\mu\nu} - \Pi \Delta^{\mu\nu}, \quad
\delta n^{\mu} \equiv - \frac{n}{\varepsilon+p} q^\mu
\ee
with
\be
u_\mu q^\mu = 0, \quad u_\mu \pi^{\mu\nu} = u_\mu \pi^{\nu\mu} = 0.
\ee
The original derivation of the IS equations is not a
systematically controlled approximation of the transport theory because it is not an expansion in some
small parameter.
In Ref.~\cite{Israel:1979wp} a quadratic {\em ansatz} for the deviation from
local equilibrium has been employed.
Nevertheless, in Ref.~\cite{Denicol:2010xn} a new method for deriving the
fluid-dynamical equations has been proposed.
In this novel approach the equation for the dissipative currents are directly
obtained from the definitions of the currents.
We note that, although these equations of motion are formally identical to the
original IS equations, the coefficients are different.
For a detailed overview of the IS theory see Ref.~\cite{Huovinen:2008te}.
Here we remark that the starting point is given by an entropy current that
includes terms up to quadratic order in dissipative quantities.
These terms are expressed - using Landau frame notation - by the same
coefficients $\ga_0,\ga_1,\gb_0,\gb_1,\gb_2,$
that encode additional transport properties.
In particular, the set $\{\beta_i\}$ describes
the relaxation times for dissipative quantities as
\be
\tau_\Pi = \zeta \beta_0 \ , \qquad  \tau_{\rm q} = \kappa_{\rm q} T \beta_1 \ , \qquad
\tau_\pi = 2\eta_{\rm s} \beta_2 \ .
\label{reltimes_IS}
\ee
The NS theory is recovered
when all these coefficients are set to zero
$\beta_0 = \beta_1 = \beta_2 = \alpha_0 = \alpha_1=0$. 

In the following we focus on  a viscous, longitudinally 
boost-invariant system with transverse translation invariance and
vanishing bulk viscosity.
As a boost-invariant system we consider a
system with a longitudinal scaling flow, $\vec v=(0, 0, z/t)$, and where all scalar
quantities are independent of the spatial rapidity, defined by 
\be
\eta= \frac{1}{2}\ln [(t+z)/(t-z)].
\ee
If the initial densities are assumed to depend on $t$ and
$z$ only through the Bjorken (longitudinal) proper time
\be\label{def:Bjtau}
\tau=\sqrt{t^2-z^2},
\ee
the expansion will evolve such that
densities remain independent of $\eta$. The $v_z$ will retain the
scaling form $v_z=z/t$~\cite{Bjorken:1982qr}. Accordingly, all vector and
tensor quantities
can be obtained from their values at
$\eta=0$ by an appropriate Lorentz boost.
Because of the symmetries of the system, i.e., longitudinal 
boost invariance, axial symmetry in the transverse plane and $\eta \to -\eta$ reflection
symmetry, the heat flow is zero everywhere.
Furthermore, only the viscous corrections to the
longitudinal, $\pi_{\rm L}$, and transverse pressure, $\pi_{\rm T}$, i.e.,
$\pi_{\rm zz}$ and
$\pi_{\rm xx}=\pi_{\rm yy}$ components of the shear stress tensor evaluated in
the local rest frame, are non-vanishing.
The choice to neglect bulk viscosity is sensible since shear viscosity is expected to
dominate at RHIC. 
The absence of heat flow as well as bulk viscosity leads to
\be
\tau_\Pi = \tau_{\rm q} = 0.
\ee
Only the relaxation time, $\tau_\pi$, enters in the equation of motion.
With these assumptions the equations of motion can be written as
\ba
\dot n + \frac{n}{\tau} &=& 0 
\label{eqn:n_bj1D}
\\
\dot \varepsilon + \frac{\varepsilon + p}{\tau} &=& 
 - \frac{\pi_{\rm L}}{\tau} 
\label{eqn:e_bj1D}
\\
\tau_\pi \dot \pi_{\rm L} 
+ \pi_{\rm L}
  \left[1 + \frac{\tau_\pi}{2\tau} 
        + \frac{\eta_{\rm s} T}{2} \dot{\left(\frac{\tau_\pi}{\eta_{\rm s} T}\right)}
  \right]
 &=&  -\frac{4\eta_{\rm s}}{3\tau} 
\label{eqn:piL_bj1D}
\\
\pi_{\rm T} &=& -\frac{\pi_{\rm L}}{2} \ ,
\label{eqn:piT_bj1D}
\ea 
where the 'dot' denotes ${\rm d}/{\rm d}\tau$.
This special case is 
a useful approximation to the early longitudinal expansion stage of 
a heavy ion collision for observables near midrapidity $\eta \approx
0$~\cite{Huovinen:2008te}.
Obviously, Eq.(\ref{eqn:n_bj1D}) describes particle conservation and can
be solved as
\be\label{eqn:n_bj1D-2}
n(\tau) = \frac{\tau_0\, n(\tau_0)}{\tau}.
\ee
In Ref.~\cite{Huovinen:2008te} these equation of motions have been studied with
the assumption of an ideal massless equation of state.
Because of this oversimplification the density equation (\ref{eqn:n_bj1D})
decouples entirely leading to two
coupled equations for the equilibrium pressure and the viscous
correction, $\pi_{\rm L}$. For two limiting scenarios, - for the scale invariant case of a constant 
shear viscosity to entropy
density ratio, $\eta_{\rm s}/s$, and for dynamics driven by a constant cross
section - analytic approximate solutions can be found.
However, we focus on a realistic equation of state and therefore the limitation
of a massless ideal system must be neglected.
We then use the results of the virial expansion for the equation of
state~\cite{Mattiello:2009fk} as well as for the shear
viscosity~\cite{Mattiello:2009db}, where, within a kinetic
approach in the ultra relativistic regime, the transport cross section as
function of the temperature has been consistently evaluated.
Therefore, we use the relaxation time $\tau_\pi$ with the
ultra relativistic assumption. As in Ref.~\cite{Israel:1979wp} we obtain
\be
\tau_\pi= \frac{3\eta_{\rm s}}{2p}.
\label{eqn:taupi}
\ee 
Note that in this way we can systematically investigate the interplay of
different effects - realistic equation of state, dissipation, dependence on
the initial condition- in the same framework.
In the following we briefly discuss these different scenarios in detail and
give the analytic solution of the equation of motion, where possible.

\subsection{Ideal Hydrodynamics}\label{Subsec:ID}
The simplest case is that the system does not suffer any dissipative
effects.
Accordingly,  we have 
\be
\tau_\pi^{\rm ID}= \eta_{\rm s}^{\rm ID}=0,
\ee
and the relevant equation of motion is 
\be
\dot \varepsilon + \frac{\varepsilon + p}{\tau} =0.
\label{eqn:e_bj1D-ID}
\ee
This automatically follows from $\pi_{\rm L} =\pi_{\rm T}=0$.
Equation (\ref{eqn:e_bj1D-ID}) is equivalent to
\be\label{eqn.IntegralofMotion}
\frac{{\rm d}}{{\rm d}\tau}(s \tau)=0,
\ee
which indicates that $s\tau$ is a conserved quantity.
This conservation law is broken if dissipative effects emerge.
Thus the entropy density evolution is given by
\be
s(\tau)=\frac{s_0\tau_0}{\tau},\qquad{\rm with }\quad s_0=s(\tau_0).
\ee
Consequently, using an EoS a solution of the equation of
motion (EoM) can be given without explicitly solving Eq.(\ref{eqn:e_bj1D-ID}).
A formal solution for the energy density evolution
$\varepsilon=\varepsilon(\tau)$ has the form
\be
\varepsilon(\tau)=\varepsilon_0\left(\frac{\tau_0}{\tau}\right)^{1+c_{0}^2},\qquad{\rm
  with } \quad c_0^2=\frac{p}{\varepsilon}\quad{\rm and }\quad \varepsilon_0=\varepsilon(\tau_0).
\ee 

This results from the separable structure of the EoM, or, more
precisely, because it is an exact differential equation.
Obviously, this formal solution holds also for the Stefan-Boltzmann limit,
where $c_0^{\rm SB}=1/3$ coinciding with the sound velocity defined by
\be
c_{\rm s}^2=\frac{\partial P}{\partial \varepsilon}=\varepsilon\frac{\partial c_0^2}{\partial\varepsilon}+\frac{P}{\varepsilon}.
\ee
This leads to the well known power law $\varepsilon^{\rm
  SB}(\tau)\propto (\tau_0/\tau)^{-4/3}$.
In general for a realistic EoS an explicit solution, $\varepsilon(\tau)$, is not possible, because the $c_0$ itself
is a function of $\varepsilon$.
However, for the QGP, the ratio pressure to energy density can be
parameterized using the phenomenological {\em ansatz}~\cite{Ejiri:2005uv}
\begin{equation}
\frac{p}{\varepsilon}=\frac{1}{3}\left(C-\frac{A}{1+B\varepsilon}\right),
\end{equation}
that provides a good fit to the lattice data in the interval
$1.3\leq\varepsilon^{1/4}/({\rm GeV/fm^3})^{1/4}\leq 6$ with
$C=0.964(5)$, $A=1.16(6)$ and $B=0.26(3){\;\rm fm^3/GeV}$~\cite{Cheng:2007jq}.
Using  this parametrization we can explicitly solve the equation of motion
(\ref{eqn:e_bj1D-ID}) as
\be
F(\varepsilon)=F_0\frac{\tau_0}{\tau}\qquad{\rm with }\quad F_0=F(\varepsilon_0).
\ee
The solving polynomial function is given by
\be
F(\varepsilon)=\varepsilon^{\frac{1}{D}}(\varepsilon+\omega)^{\frac{3}{C+3}-\frac{1}{D}}
\ee
with the constants,
\ba
D&=&\frac{C}{3}+1-A\\
\omega&=&1-\frac{3A}{C+3}=\frac{D}{B(1+C/3)}.
\ea
naturally, setting $A=0$ and $C=1$ - and $D=4/3$ and $\omega=1$ - the
evolution in the SB-limit is recovered.

\subsection{Navier-Stokes}\label{Subsec:NS}
As mentioned previously, the simplest way to introduce dissipative effects is
considering the linear Navier-Stokes equations. In this approximation all
relaxation times are vanishing and consequently the equations for the energy density
evolution and the viscous correction, $\pi_{\rm L}$, completely decouple
\ba\label{eqn:EoM_NS}
\dot \varepsilon + \frac{\varepsilon + p}{\tau}&=&\frac{4}{3}\frac{\eta_{\rm
    s}}{\tau^2},\\
\pi_L &=& -\frac{4\eta_s}{3\tau}.
\label{eqn:piL_NS}
\ea
The NS equation of motion (\ref{eqn:EoM_NS}) can be reduced to an
exact differential equation.
Formally, using the same notation as before, we can write
\be
\varepsilon(\tau)=\varepsilon_0\left(\frac{\tau_0}{\tau}\right)^{1+c_{0}^2}H_{\rm
NS}(\eta_{\rm s}),
\ee
where the dissipative correction, $H_{\rm NS}$, is given by 
\be\label{eqn:H_NS}
H_{\rm NS}(\eta_{\rm s})=\exp\left\{-\frac{4}{3}\int_{\tau_0}^\tau {\rm
  d}\tau' \frac{\partial \eta_{\rm s}}{\partial\varepsilon}\frac{1}{\tau'^2}\right\},
\ee
Eq. (\ref{eqn:H_NS}) corresponds to the integration factor needed to transform the NS equation
motion in to an exact differential equation.
This structure can be found as well in Ref.~\cite{Cheng:2001dz},
where, by assuming for the evolution of the viscosity 
\be\label{eqn:viskNL}
\eta_{\rm s}^{\rm nl}=C_{\rm NS}\;\varepsilon\tau+C_{\rm nl}\;\varepsilon/\tau,
\ee
non local corrections governed by the parameter, $C_{\rm nl}$, have been
included.
With this analytic form for the viscosity the dissipative correction given by
$H_{\rm NS}$ can be easily calculated, and one finds
\ba
\varepsilon^{\rm nl}(\tau)&=&\varepsilon(\tau_0)\left(\frac{\tau_0}{\tau}\right)^{(1+c_0^2-(4/3)C_{\rm NS})}
H_{\rm NS}^{\rm nl}(\eta_{\rm s}^{\rm nl})\nonumber\\
&=&\varepsilon(\tau_0)\left(\frac{\tau_0}{\tau}\right)^{(1+c_0^2-(4/3)C_{\rm NS})}
\exp\left\{\frac{2C_{\rm nl}}{3}\left(\frac{1}{\tau_0^2}-\frac{1}{\tau^2}\right)
\right\},\label{eqn:non-localepsilon}
\ea
as shown in~\cite{Cheng:2001dz} for massless non-interacting particles,
i.e., setting $c_0^2=1/3$.
This assumption for the viscosity means that an additional coupled
equation for the time evolution of $\eta_{\rm s}$ has been implicitly added in the
investigation.
This is not our strategy, because we have completely determined the shear
viscosity in a systematic way, and we can also calculate the time
evolution without further assumptions.
In fact, the EoS and the knowledge of the dependence of the viscosity on
the thermodynamics quantities,.i.e., on the energy density, $\varepsilon$~\cite{Mattiello:2009fk,Mattiello:2009db},
allows us to find a coupled system of equations for
the NS scenario as well as in the general case. 
Therefore, by finding a parametrization of $\eta_{\rm
  s}=\eta_{\rm s}(\varepsilon)$,
we are able to give an analytic solution calculating $H_{\rm NS}^{\rm
  Vir}(\eta_{\rm s})$.
We note in passing that in this way the equation for the energy density and
for the viscous correction for the pressure decouple.
However, we prefer to directly solve the equations of motion without introducing any parametrization for the viscosity.


\section{Entropy production}\label{Sec:EntrProd}

Now we define the basic observable investigated in this study, and consider
its different evolution in the different scenarios.
In the rest of the paper, 
the subscript '0' refers to the value of
quantities at the initial time, $\tau_0$.

The most natural quantitative measure of dissipative effects is the
entropy production.
An often used quantity is the entropy per unit rapidity
\be
\frac{{\rm d}S}{{\rm d}y} = \tau \int \dd^2 x_{\rm T}\,s =\tau A_{\rm T} s,
\label{dSdeta}
\ee
where $A_{\rm T}$ is the transverse area of the system.
This quantity highlights several aspects of the QGP because the
final entropy production can be measured~\cite{Pal:2003rz} and, additionally, it is
sensitive to the dissipative properties of the system.
In ideal hydrodynamics, the entropy per unit
rapidity is evidently a constant of the evolution. 
By the experimental measurement of the entropy production at the final time
$\tau_{\rm f}$ one can conclude the initial condition as 
\be
\frac{{\rm d}S^{\rm ID}}{{\rm d}y}(\tau_{\rm f}) = \frac{{\rm d}S^{\rm ID}}{{\rm d}y}(\tau_0) = \tau_0 A_{\rm T} s_0.
\label{dSdeta-ID}
\ee
In the NS scenario and in the IS regime, dissipative effects determine an enhancement of the entropy
production.
This viscous correction depends of course on the initial conditions,
specifically on the formation
time and initial energy density as usual, but in general also on the initial
pressure anisotropy.
Therefore, the most natural choice for the initial conditions are the initial
entropy density, $s_0$, and the initial ratio of viscous longitudinal shear and
equilibrium pressure, $\xi_0$, defined as
\be
\xi_0 \equiv \frac{{\pi_{\rm L}}_0}{p_0}.
\ee
A useful equivalent measure is the pressure anisotropy coefficient
\be
R_0\equiv \frac{{p_{\rm L}}_0}{{p_{\rm T}}_0}  = \frac{1+\xi_0}{1-\xi_0/2} \ , 
\ee
which is the ratio of the transverse and longitudinal pressures
${p_{\rm T}}_0 \equiv p - {\pi_{\rm L}}_0/2$, ${p_{\rm L}}_0 \equiv p_0 + {\pi_{\rm L}}_0$.
Evidently, $\xi_0=0$ indicates central collisions, where pressure anisotropy
can not develop, whereas peripheral collisions lead to larger $\vert\xi_0\vert$.

We have shown in Section~\ref{Subsec:ID} that in ideal hydrodynamics the
viscous corrections to the pressure
always vanish. In this scenario the anisotropy is unity and 
\be
R_0^{\rm ID} = 1\qquad{\rm and}\quad \xi_0^{\rm ID} =0.
\ee
are automatically fulfilled.

As explained in Section~\ref{Subsec:NS}, in the NS approximation the evolution of the energy density
and of the viscosity are uniquely determinated by
Eq.(\ref{eqn:EoM_NS}). Therefore, the initial anisotropy is automatically fixed
by the initial condition over the entropy density $s_0$ and any dependence on
$\xi_0$ (or $R_0$) vanishes.
In IS the dependence on the initial pressure anisotropy emerges and thus
in this scenario we need it as a second initial condition.
Additionally, Eq.(\ref{dSdeta}) shows that the entropy production is proper time dependent.
Therefore, the final proper time $\tau_{\rm f}$ enters as an important parameter to
describe the experimental data.

\section{Application to 130 GeV Au + Au collisions}\label{Sec:Results}

As mentioned in the previous section the entropy production allows -- in
principle -- to quantify the dissipative effects. Furthermore, experimental
results are available for Au + Au collision at RHIC at $\sqrt{s_{\rm NN}}=130$
GeV~\cite{Pal:2003rz}, where the final entropy production per unit rapidity
is 4501 with an uncertainty of about $10 \%$.
In the following we attempt to categorize which (dissipative)
effects and initial conditions are compatible with these experimental
findings.
As in Ref.~\cite{Pal:2003rz} we compare the experimental value to  the
hydrodynamic ${\rm d}S/{\rm d}y$ as a function of the initial energy density, $\varepsilon_0$.
We emphasize that $\varepsilon_0$ is strongly dependent
on the choice of the formation time, $\tau_0$. In Ref.~\cite{Pal:2003rz} the
authors use a conservative formation time, $\tau_0=1$ fm/c. 

\begin{figure}[ht]
\centerline{\epsfig{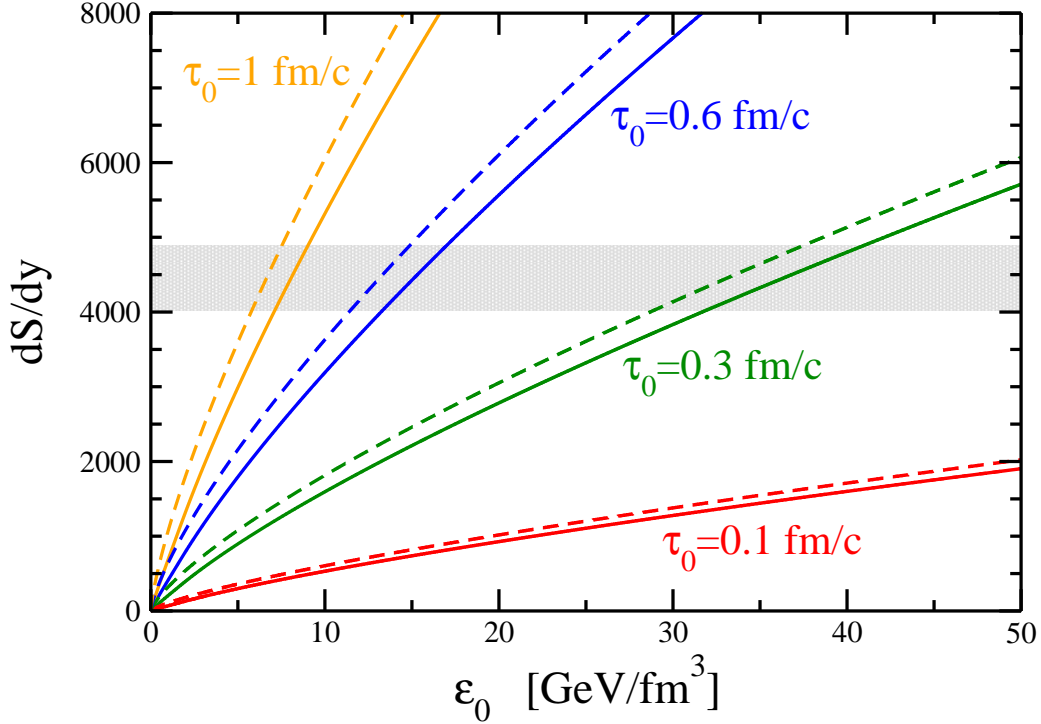}}
\caption{(Color online) The entropy per unit rapidity is displayed as a function of the
  initial energy density $\varepsilon_0$ for the QGP using a realistic
  EoS~\cite{Mattiello:2009fk} (solid lines) in comparison to the SB limit
  (dashed lines) for different formation times, $\tau_0=0.1,\,0.3,\,0.6,\,1$ fm/c (red, green, blue and
  orange lines). The horizontal band shows the final-state entropy extracted from
experiment. Hydrodynamic simulations typically have energy densities between
10 and 15 GeV/fm$^3$ and formation times of about 0.6 fm/c.
}
\label{fig:dSdy_Eps0_SBuND}
\end{figure}
Our aim is the investigation of the role of all initial conditions and therefore the
explicit dependence on the formation time has to be considered.
As a first step, we focus on the importance to use a realistic EoS,
in order to clarify the amount and the behavior of the effects that a realistic EoS generates in the experimental observables.
In Fig.~\ref{fig:dSdy_Eps0_SBuND} the entropy per unit rapidity is shown as a function of the
  initial energy density, $\varepsilon_0$, for the QGP using the realistic
  EoS of~\cite{Mattiello:2009fk} (solid lines ) 
for various formation times.
In following we refer to such calculation as non dissipative scenario (ND).
For comparison, we indicate by the dashed line the SB-limit for the same
initialization times using the same color coding.  The horizontal band shows the final-state entropy extracted from
experiment. 
Evidently, too fast thermalization times, $\tau_0=0.1$ fm/c and $\tau_0=0.3$
fm/c, lead to an unreasonably high initial energy density. 
Hydrodynamic simulations typically use formation times about 0.6 fm/c, which
corresponds to the blue lines of Fig.~\ref{fig:dSdy_Eps0_SBuND}.
The corresponding results for the entropy in our calculation suggests energy densities between
10 and 15 GeV/fm$^3$, which is in agreement with ideal hydrodynamic
simulations. We remark here that for this typical hydrodynamic formation time
as well as for the conservative one, $\tau_0=1$ fm/c, used in several
evaluations (extrapolations) of measured experimental data, the difference
between SB and realistic EoS is not negligible.
Therefore, the precise determination of the initial conditions is necessary
for a good description of the evolution of the system. 
\begin{figure}[t]
\centerline{\epsfig{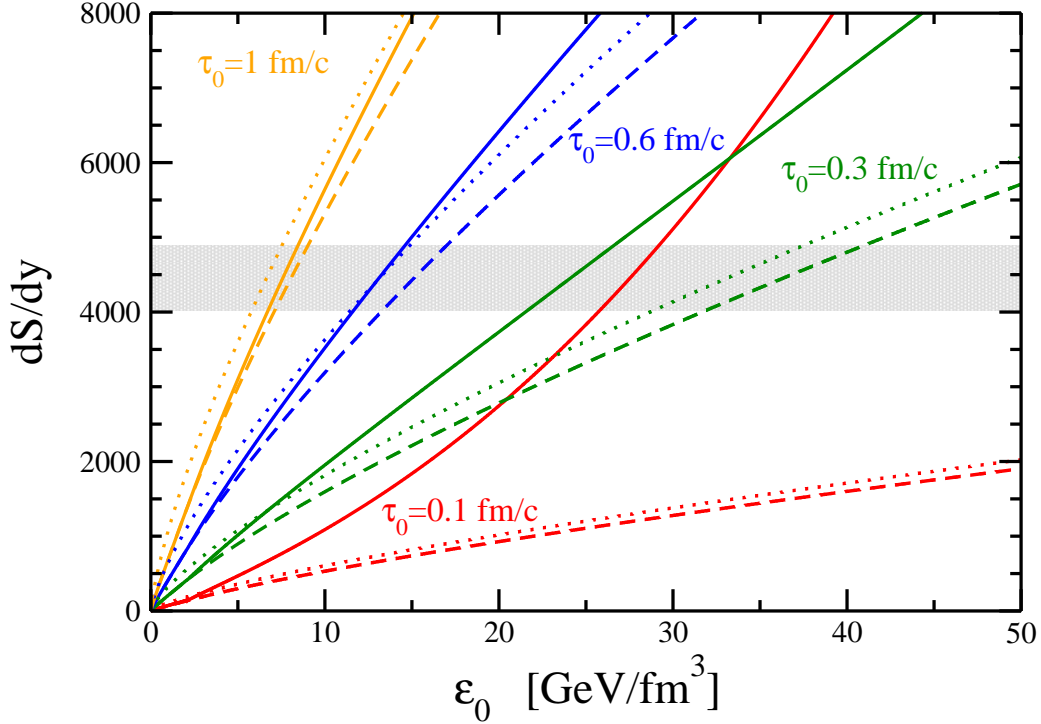}}
\caption{(Color online) The entropy per unit rapidity is picted as a function of the
  initial energy density, $\varepsilon_0$, for a dissipative realistic
  QGP~\cite{Mattiello:2009fk,Mattiello:2009db}
within the NS approximation (solid lines) in comparison to
  the ND results
  (dashed lines) and the SB limit (pointed line) for different formation times, $\tau_0=0.1,\,0.3,\,0.6,\,1$ fm/c (red, green, blue and
  orange lines). The horizontal band shows the final-state entropy extracted from
experiment. }
\label{fig:dSdy_Eps0_NSuND}
\end{figure}
In this context, an implementation of the viscosity effect
have a larger impact.

Because of the dissipative effects in
the QGP, the entropy density per unit rapidity is an increasing function of the
proper time. 
Following~\cite{Song:2008si} we assume that the final entropy is mostly produced in the early phases of the collisions,
and then we can neglect the contribution of the hadronic phase.
Therefore the final time, $\tau_{\rm f}$, is automatically fixed, namely as the time when the energy density of the system is equal to the critical energy density,
\be
\varepsilon(\tau_{\rm f})=\varepsilon(T_{\rm c}).
\ee
Because in the NS approximation the initial viscosity of the system
is completely determined by the initial energy density,
we can directly compare the results for the entropy per unit rapidity
for the non dissipative system shown before with the ones from the NS equation
of motions.
Using the same color coding as before, we show in
Fig.~\ref{fig:dSdy_Eps0_NSuND} the entropy per unit rapidity as a function of
the initial energy density $\varepsilon_0$ for the dissipative
QGP~\cite{Mattiello:2009fk,Mattiello:2009db} within the NS
approximation (solid lines) for various formation times in comparison with the
non dissipative and SB calculations.
For all formation times one observes an enhancement of the entropy
production by the NS viscous corrections in comparison to the ND results.
For $\tau_0=1$ fm/c and $\tau_0=0.6$ fm/c we find moderate corrections, that seem to recover the results of the SB ideal fluid
shown in Fig.~\ref{fig:dSdy_Eps0_SBuND}.
In particular, for $\tau_0=0.6$ fm/c, the effect of a realistic equation of
state and the linear (NS) viscous terms compensate also quantitatively each other. For the faster thermalization scenarios, $\tau=0.3$
fm/c and $\tau=0.1$ fm/c, the enhancement is very significant and allows to
describe the experimental finding for the entropy per unit of rapidity using
smaller -- but nevertheless high -- $\varepsilon_0$ of about 20 GeV/fm$^3$ and 25 GeV/fm$^3$
respectively.
The formation time, $\tau_0$, indicates an atypical behavior, not
only because of the pronounced enhancement of the entropy production, but
primarily because of the change from a concave to a convex function in
${\rm d}S/{\rm d}y(\varepsilon_0)$.
This can be a hint that results in NS approximation with small formation
time have to be questioned.
Deciding this question, one has to consider the full IS equations of motion
for the entropy production.

As explained in Section~\ref{Sec:EntrProd} the initial conditions for the
IS equation of motion are not completely fixed by the initial energy density
of the system because of the additional explicit anisotropy of the system.
Therefore, one has a family of solutions labeled by the anisotropy coefficient, $\xi_0$, as additional parameter.
For each formation time, we calculate, for different values of
the initial anisotropy, the entropy production at $T=T_{\rm c}$ as function of
the initial energy density.
We use $\xi_0=0$, $\xi_0=-0.5$, and $\xi_0=-1$, which correspond to the values for
the anisotropy coefficient of $R_0=1$, $R_0=2/3$, and $R_0=0$, respectively.
We prefer to concentrate on two limiting cases, $\tau_0=0.1$ fm/c
and $\tau_0=1$ fm/c.

\begin{figure}[t]
\centerline{\epsfig{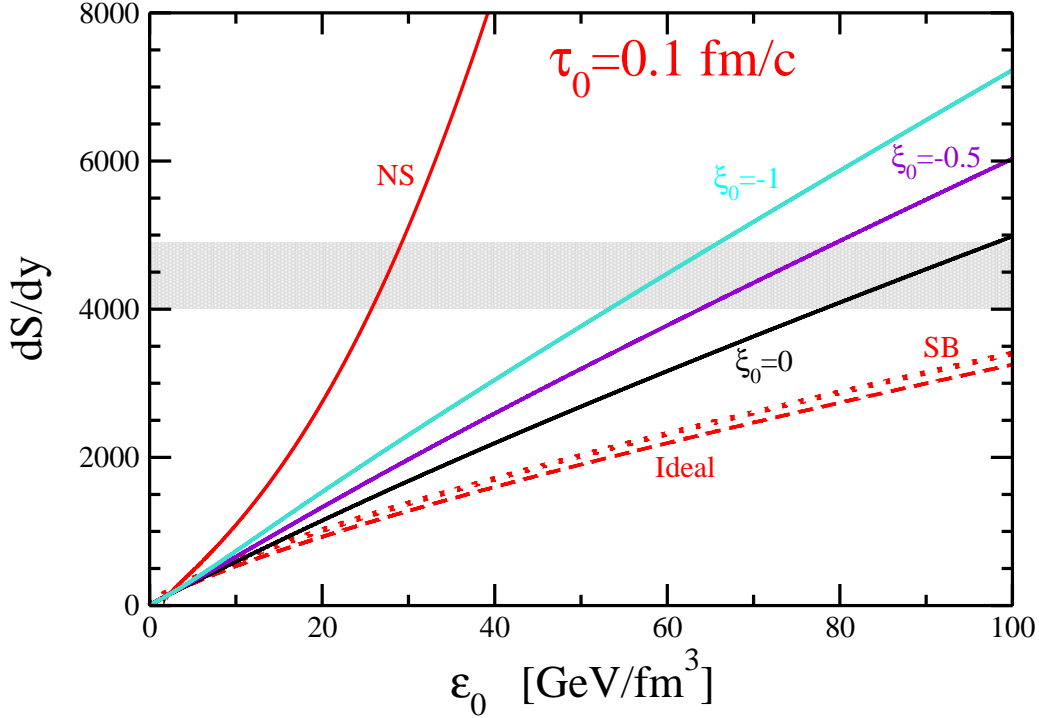}}
\caption{(Color online) The entropy per unit rapidity is depicted as a function of the
  initial energy density, $\varepsilon_0$, for a dissipative realistic
  QGP~\cite{Mattiello:2009fk,Mattiello:2009db} at $\tau_0=0.1$ fm/c
within the IS approach for different values of the initial pressure anisotropy,
$\xi_0=0,-1,-2,-3$ (black, turquoise, violet and brown solid line
respectively) in comparison to the NS approximation (red solid line), the
ND scenario (dashed red line) and the SB limit (dotted red line). The horizontal band shows the final-state entropy extracted from
experiment.}
\label{fig:dSdy_Eps0_ISt_eq0_10uNS}
\end{figure}

In Fig.~\ref{fig:dSdy_Eps0_ISt_eq0_10uNS} we show the entropy per unit rapidity as a function of
the initial energy density $\varepsilon_0$ with the formation time
$\tau_0=0.1$ fm/c  within the IS
approach. The different values of the initial pressure anisotropy
$\xi_0=0,-0.5,-1$ are displayed as black, violet and turquoise lines,
respectively.
\begin{figure}[t]
\centerline{\epsfig{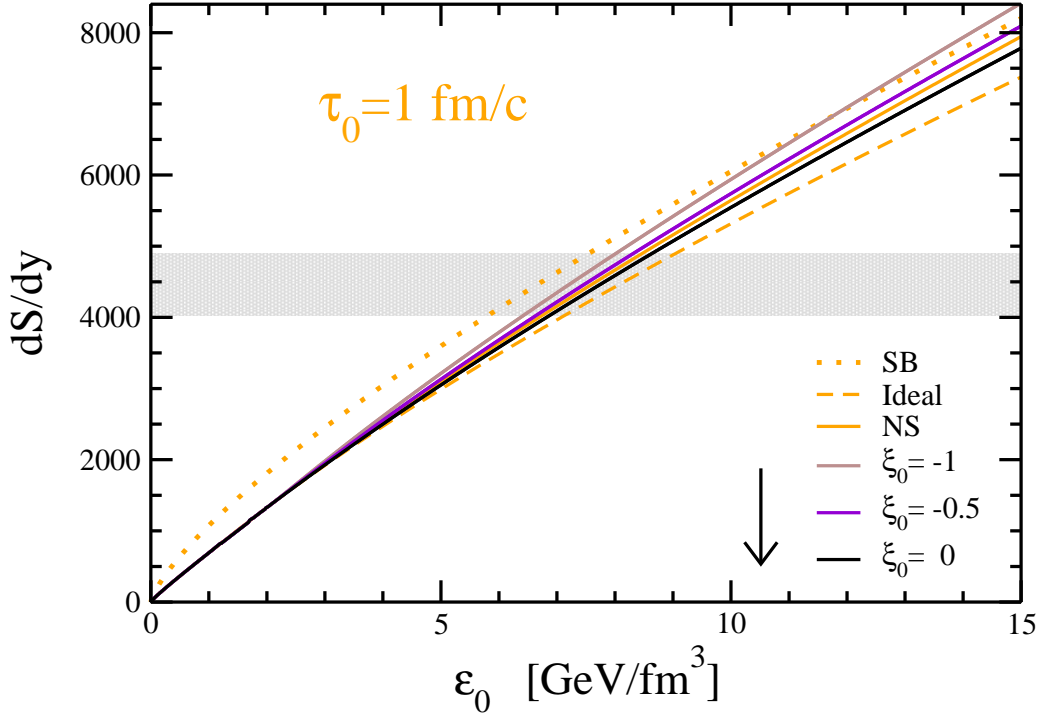}}
\caption{(Color online) The entropy per unit rapidity is displayed as a function of the
  initial energy density, $\varepsilon_0$, for a dissipative realistic
  QGP~\cite{Mattiello:2009fk,Mattiello:2009db} at $\tau_0=1$ fm/c
within the IS approach for different values of the initial pressure anisotropy,
$\xi_0=0,-1,-2,-3$ (black, turquoise, violet and brown solid line
respectively) in comparison to the NS approximation (orange solid line) and the
ND scenario (dashed orange line) and the SB limit (dotted orange line).The horizontal band shows the final-state entropy extracted from
experiment.}
\label{fig:dSdy_Eps0_ISt_eq1_00uNS}
\end{figure}
For comparison we added the corresponding results within the NS
approximation (red solid line), the ND scenario (dashed red line) and the SB
limit (dotted red line).
The experimental result is located within the horizontal band in Fig.~\ref{fig:dSdy_Eps0_ISt_eq0_10uNS}. 
Already at vanishing anisotropy the deviation from the previous
results is evident in the whole energy range.
the difference to the non dissipative scenario are small for low initial
energy density, $\varepsilon\geq 10$ GeV/fm$^3$.
By increasing the anisotropy of the system, the entropy production also increases.
For all values of $\xi_0$ we observe a linear relation between ${\rm
  d}S/{\rm d}y$ and $\varepsilon_0$.
Nevertheless, we reject such a fast formation time for two reasons: not only does the
NS formulation not allow $\tau_0=0.1$ fm/c, but also at extremely high initial
anisotropy $\xi_0=-1$, that corresponds to the limiting case with the tranverse
pressure only, high values of the initial energy density are needed to
reproduce the experimental entropy production. That makes this scenario unlikely. 
The contrary situation is given by a conservative formation time, $\tau_0=1$ fm/c.

In this case Fig.~\ref{fig:dSdy_Eps0_ISt_eq1_00uNS} shows the results for the entropy per unit rapidity as a function of
the initial energy density $\varepsilon_0$ within the IS
approach for the same values of $\xi_0$ as before.
Again, we compare to the corresponding results in the NS
approximation (orange solid line), the ND scenario (dashed orange line) and
the SB limit (dotted orange line).
We note the same qualitative behavior as for the small formation time (Fig.~\ref{fig:dSdy_Eps0_ISt_eq0_10uNS}):
by increasing the anisotropy of the system, the entropy production also
increases. In particular, we find an initial energy
density between $\varepsilon_0\approx 7-9\;{\rm GeV/fm}^3$, that is larger than
experimental estimation in Ref.~\cite{Adcox:2001ry} of about
$\varepsilon_0\approx 4-5 \;{\rm GeV/fm}^3$.
This is not surprising, because the experimental measure is based
on very simplified and strong assumptions, that
do not give a realistic description of the system.
This suggests that the estimation procedures have to be improved because of the
effects of the final proper time $\tau_{\rm f}$, the viscosity and the
anisotropy can not be neglected.
To be more quantitative we can extract from the experimental estimation of the
final entropy the initial energy density for different formation times in the
different scenarios.
In this way we can compare the (typical experimental) values, $\varepsilon^{\rm
  SB}_0$, obtained using the ideal Bjorken expansion, i.e., the SB limit, with the values
$\varepsilon^{\rm ND}_0$, $\varepsilon^{\rm NS}_0$ and $\varepsilon^{\rm
  IS}_0$ in the ND scenario, in the NS and IS theory respectively.
Evidently, for the IS calculation, non only the parametric dependence on the
formation time $\tau_0$, but also on the initial pressure anisotropy $\xi_0$
has been considered.
The results are summarized in Tab.~\ref{Tab:Dev1} and Tab~\ref{Tab:Dev2}, where the deviation from the
standard SB estimation of the ND, NS and IS calculation is listed.
\begin{table}
\caption{Results for the deviation from the SB estimation of the initial energy
  density in the ND and NS scenario}\vspace*{0.5cm}
\centering
\begin{tabular}{cccc}
\hline\hline\\[-2ex]
$\tau_0$ & $\varepsilon^{\rm SB}_0$&$\varepsilon^{\rm ND}_0/\varepsilon^{\rm SB}_0$&$\varepsilon^{\rm NS}_0/\varepsilon^{\rm SB}_0$\\
${\rm [fm/c]}$&  ${\rm [GeV/fm^3}]$ &&\\[0.5ex]
\hline\\[-2ex]
0.1&143.10&1.05&0.20\\
0.3&33.07&1.10&0.73\\
0.6&13.12&1.15&0.99\\
1.0&6.64&1.21&1.14\\[1ex]
\hline
\end{tabular}
\label{Tab:Dev1}
\end{table}
\begin{table}
\caption{Results for deviation from the SB estimation of the initial energy
  density in the IS scenario}\vspace*{0.5cm}
\centering
\begin{tabular}{ccccc}
\hline\hline\\[-2ex]
$\tau_0$ & $\varepsilon^{\rm SB}_0$&$\varepsilon^{\rm IS}_0/\varepsilon^{\rm SB}_0$,&$\varepsilon^{\rm IS}_0/\varepsilon^{\rm SB}_0$,&$\varepsilon^{\rm IS}_0/\varepsilon^{\rm SB}_0$,\\
${\rm [fm/c]}$&  ${\rm [GeV/fm^3}]$ &$\xi_0=0$&$\xi_0=-0.5$&$\xi_0=-1$\\[0.5ex]
\hline\\[-2ex]
0.1&143.10&0.62&0.50&0.41\\
0.33&3.07&0.9&0.80&0.72\\
0.6&13.12&1.02&0.99&0.93\\
1.0&6.64&1.16&1.12&1.08\\[1ex]
\hline
\end{tabular}
\label{Tab:Dev2}
\end{table} 
For small formation times ($\tau_0=0.1$ fm/c and
$\tau_0=0.3$) the effects of the relativistic equations are almost
negligible, i.e. lower then $10\%$.
However, dissipative effects are to be included, because the viscosity leads
to an deviation between $30\%$ and $80\%$ in the NS scenario and between
$40\%$ and $60\%$ in the IS scenario.
Only for central collisions ($\xi_0=0$) the deviation in the formation time
$\tau_0=0.3$ fm/c is maybe reasonable ($10 \%$), although, in this case, the
assumption leads to an unlikely large initial energy density.
Therefore, such small formations time have to be rejected.
For $\tau_0=0.6$ fm/c, often used in hydrodynamical calculations, we note that
the viscous effects completely compensate the correction of
the realistic EoS in the NS scenario and in the IS theory for non peripheral collisions.
Thus, in this case, using the Bjorken expansion can be justified.
For the conservative formation time, $\tau_0=1$ fm/c, the effects of the
realistic EoS are more important, about $20\%$, and can not be compensated by the inclusion of NS
viscous corrections. Therefore, it is not surprising, that also in central
collisions in the IS scenario the deviation is also sizable.
For more peripheral collisions the SB limit seems to lead to a more or
less satisfactory approximation (about $8\%$) for the IS calculations.
Nevertheless, from this discussion it is evident, that the use of the Bjorken
expansion for the evaluation of the initial energy density, without
considerations of interplay between formation time, viscosity, and the centrality of the
collisions can be very questionable.
Additionally, the quite satisfactory agreement between IS hydrodynamics
calculations and the transport
ones within the cascade BAMPS for small $\eta_{\rm s}$~\cite{Bouras:2009zz,El:2010mt} also remarks
the validity of the IS scenario.

\section{Conclusions}\label{Sec:Concl}

In this work we have discussed the importance of the initial condition
and the role of the viscosity for the evolution of the fireball of heavy-ion collisions.
We have used a realistic EoS derived within a virial
expansion~\cite{Mattiello:2009fk}, that is in line with recent
three-flavor lattice QCD data~\cite{Cheng:2007jq}.
The shear viscosity has been consistently calculated within this
formalism using a kinetic approach in the ultra-relativistic regime with an
explicit and systematic evaluation of the transport cross section as a function
of temperature~\cite{Mattiello:2009db}.
We explicitly considered different scenarios: ideal hydrodynamic, dissipative
effects in the Navier-Stokes as well as in the Israel-Stewart formalism, from conservative to very
fast equilibration dynamics.
We choose the parameter of these studies in order to describe the experimental
findings of the entropy production for Au + Au collision at RHIC at
$\sqrt{s_{\rm NN}}=130$ GeV.
The assumption of a fast equilibration would require unreasonably high values
of the initial energy density.
With the conservative ($\tau_0=1$ fm/c) and the typical hydrodynamical
($\tau_0=0.6$ fm/c) formation times the initial energy density needed to
reproduce the final entropy is more in line with the experimental estimations.
In these scenarios, the interplay between effects of the viscosity and of the
realistic EoS can not be neglected in the reconstruction of the initial
state from the experimental data.
Additionally, centrality dependence have to be considered, because different
impact parameters lead to different initial anisotropy values, which
modify sizeably the estimation of the initial energy density. 
In conclusion, our investigation shows that from the experimental final
entropy it is very hard to
derive unambiguous information about the initial conditions and/or the
evolution of the system.
The choice of the formation time, the viscosity, and the initial anisotropy
are interplaying aspects which have to be included in the estimation of the
initial condition.
Therefore, we suggest that the easy extraction rule used should be revised for
a a better estimation of the uncertainty in the measurement.
Of course, this improvement is model dependent, but we note that also the usual
evaluation using the Bjorken expansion is  a model calculation.

Furthermore, the solution of the IS equation with a realistic equation of state and with
the inclusion of viscosity leads
to further interesting applications in the description of the heavy-ion
collisions, as (semi)analytic solution of the hydrodynamical equation in
accord of Ref.~\cite{Csanad:2009wc}.
Additionally, a systematic calculation of the bulk viscosity within the virial expansion and the
kinetic theory can be implemented to consider all remaining
dissipative effects in the QGP in a systematic way.

Acknowledgment: 
I thank  H. van
Hees, P. Huovinen, D. Rischke, P. Romatschke, and Stefan Strauss for useful discussions and suggestions.
This work is supported by Deutsche Forschungsgemeinschaft and by the Helmholtz International Center for FAIR  
within the LOEWE program of the State of Hesse.

\bibliographystyle{elsarticle-num-names}
\bibliography{literature}

\end{document}